\documentclass{ws-mpla}
\usepackage[super]{cite}
\usepackage{graphicx}
\usepackage{colortbl}
\newcommand {\pd}{\partial}
\begin{document}

\markboth{Authors' Names}
{Instructions for Typing Manuscripts (Paper's Title)}

\catchline{}{}{}{}{}

\title{De Sitter and power-law solutions in some models of modified gravity}

\author{Yi Zhong}

\address{Institute of Theoretical Physics,
            Lanzhou University, \\Lanzhou 730000,
            People's Republic of China\\
            Institute of Space Science, ICE-CSIC and IEEC,\\ Campus UAB,
    Carrer de Can Magrans s/n, 08193 Bellaterra, Barcelona, Spain
	zhongy13@lzu.edu.cn}

\author{Emilio Elizalde}

\address{Institute of Space Science, ICE-CSIC and IEEC,\\ Campus UAB,
    Carrer de Can Magrans s/n, 08193 Bellaterra, Barcelona, Spain\\
    elizalde@ieec.uab.es}

\maketitle

\pub{Received (Day Month Year)}{Revised (Day Month Year)}

\begin{abstract}
Inspired by some recent works of Lovelock Brans-Dicke gravity and mimetic gravity, cosmology solutions in extensions of these two modified gravities are investigated. A non-local term is added to the Lovelock Brans-Dicke action and Gauss-Bonnet terms to the mimetic action,correspondingly. De Sitter and power scale factor solutions are then obtained in both theories. They can provide natural new approaches to a more accurate description of the unverse evolution.

\end{abstract}

\ccode{PACS Nos.: 4.50.Kd.}


\section{Introduction}
In current cosmology two intriguing problems are apparent, namely the dark energy, and the dark matter problems. The first of them can be phrased as to  what are the specific mechanics of the accelerating expansion of the universe.  Also, the relation that may exist between the present acceleration and the very extreme one that took place during inflation is being discussed. The scenarios to explain these two accelerations are usually quite similar and, as a result, several competing scenarios, valid at the present level of accuracy of the astronomical data, have been proposed in the literature. It seems now clear that the gravitational action may contain extra terms (as compared to general relativity, GR) which became relevant only recently, with an associated significant decrease of the universe curvature. Therefore. a rather natural and economic solution to these problems is provided by modified gravity. For general reviews in modified gravity, see \cite{NOJIRI_2007,Nojiri_2011}. Various approaches of modified gravities have been applied to generate the accelerating expansion.
It is shown that by adding ln R or Gauss-Bonnet terms one can generate the necessary cosmic acceleration \cite{Nojiri_2004,Nojiri_2005,Cognola_2006}.
In Ref. \cite{Elizalde_2004} the  dark energy and the cosmic speed-up in scalar-tensor theory  have been  studied.
In  Ref. \cite{Cognola_2008} viable modified f(R) gravities describing inflation and the onset of accelerated expansion  have been  studied.
General f(R) gravity at the one-loop level in a de Sitter universe was investigated in Ref. \cite{Cognola_2005}, extending a similar program
developed for the case of pure Einstein gravity.  Using generalized zeta-functions regularization they could get the one-loop effective action and then study the possibility
of stabilization of the de Sitter background by quantum effects and, such approach hints to a possible way of solving the cosmological constant problem. Hence, the study of
one-loop f(R) gravity seems a natural step to do.

Most of the models being considered are local theories. But it should be remarked that quantum gravity together with stringy considerations actually lead to a low-energy effective action containing non-local terms, in addition to the ordinary, local ones. It has been shown by different authors that these non-local terms ccould be responsible for both inflation and the current cosmic acceleration (see, e.g., \cite{PhysRevLett.99.111301,Nojiri2008821,Capozziello_2010}). On the other hand, a convincing motivation for the extension of Brans-Dicke (BD) gravity has been recently given in \cite{Tian:2015vda}, namely that alternative and modified gravities actually encode possible ways to go beyond Lovelock's theorem and its necessary conditions, which limit the second-order field equation in four dimensions to essentially Einstein's equation supplemented by the cosmological constant. In this way, similarly to what happened with the introduction of extended GR theories, extensions of BD gravity naturally come into play. To this we here add the possibility, as just discussed above, of the theory being non-local and obtain cosmological solutions for a BD non-local model. We will first realize the de Sitter solutions, corresponding to the inflation and accelerating expanding period; and then the power scale factor cosmology, corresponding to a perfect fluid.

Recently, a new type of modified gravity has been discussed where the metric $g_{\mu\nu}$ is defined as:
\begin{eqnarray}
    g_{\mu\nu}=\hat{g}^{\rho\sigma} \pd_{\rho}\phi\pd_{\sigma}\phi \hat{g}_{\mu\nu},
    \label{memetic metric}
\end{eqnarray}
$\hat{g}_{\mu\nu}$ being an auxiliary metric  and $\phi$  a scalar field. In this method, the conformal degree of
freedom is isolated in a covariant way. It is shown that this theory can mimic cold dark matter
\cite{Chamseddine_2013}. Cosmology in mimetic gravity was considered in Refs. \cite{Chamseddine_2014,Odintsov_2015,Odintsov_2016,Nojiri_2016}. and was extended into mimetic
$f(R)$ gravity in \cite{Nojiri_2014}. Inflation, dark energy and other topics have been also investigated in these theories \cite{Nojiri_2014,Deruelle_2014,Momeni_2014,Leon_2015,Momeni_2015,Raza_2015}. In the present paper we extend this theory into mimetic Gauss-Bonnet theory, i.e., we introduce the Gauss-Bonnet term into the action and  study the cosmological solutions.
The role of the Gauss-Bonnet term in view of cosmological solutions is discussed in Ref. \cite{Capozziello_2015,Myrzakulov_2015}.
 Mimetic GB gravity has been introduced in Ref. \cite{10.1088/0264-9381/32/18/185007}. Particularly, we shall investigate the de Sitter and power scale factor solutions.

This paper is organised as follows: In Sec. II, we study de Sitter and power-law solutions in
 BD non-local theory. In Sec. III we investigate the de Sitter and power-law solutions in the other case of  mimetic Gauss-Bonnet theory. The paper ends up with the conclusions.

\section{Non-local Brans-Dicke gravity}
 As explained in the Introduction, non-local gravity can be generalized further to non-local Brans-Dicke (BD) gravity \cite{Tian:2015vda}. The corresponding action of the system is
    \begin{eqnarray}
        S_{\text{NBD}}=\int d^4x\sqrt{-g} L_{\text{NBD}} + S_M,
        \label{action nbd1}
    \end{eqnarray}
where
    \begin{eqnarray}
        L_{\text{NBD}}=\frac{1}{2\kappa^2}
        \left\{ \phi R
        \left[1+f(\frac{1}{\Box}R)\right]
        -\frac{w}{\phi } \nabla _{\mu }\phi \nabla ^{\mu }\phi \right\}.
        \label{lagrangian nbd}
    \end{eqnarray}
Here $f$ is a function, $\frac{1}{\Box}R$  the non-local term, and $\Box$ is the d'Alembertian for the scalar
field. Introducing two scalar fields, $\eta$ and $\xi$, the action (\ref{action nbd1})
can be written as \cite{Nojiri2008821}
    \begin{eqnarray}
        L_{\text{NBD}} &=& \frac{1}{2\kappa^2} \left\{ \phi R [f (\eta) +1]+\xi(\Box\eta-R)
        -\frac{w}{\phi }\nabla _{\mu }\phi \nabla ^{\mu }\phi \right\}  \nonumber\\
        &=& \frac{1}{2\kappa^2}
        \left\{\phi R [f(\eta) +1]-\text{$\xi $R}-\partial_\mu \xi \partial ^{\mu }\eta -\frac{w}{\phi }\nabla _{\mu }\phi \nabla ^{\mu }\phi \right\}.
    \end{eqnarray}
Variations with respect to $\phi,\eta, \xi, g_{\mu\nu}$ give, respectively,
    \begin{eqnarray}
        \frac{2w}{\phi}\Box\phi+R+f(\eta)R-\frac{w}{\phi^2}(\partial\phi)^2=0
        \label{equation phi1}\\
        \phi R f'(\eta) + \Box \xi = 0
        \label{equation eta1}\\
        \Box \eta =R
        \label{equation xi1}\\
        2\kappa^2 \frac{{\delta {L_{{\rm{NBD}}}}}}{{\delta {g^{\mu \nu }}}} -
        \kappa^2 {g_{\mu \nu }}{L_{{\rm{NBD}}}} = \kappa^2 {T_{\mu \nu }},
        \label{gravitational equation1}
    \end{eqnarray}
where the term $2\kappa^2 \frac{{\delta {L_{{\rm{NBD}}}}}}{{\delta {g^{\mu \nu }}}}$ reads
    \begin{eqnarray}
        2\kappa^2 \frac{{\delta {L_{{\rm{NBD}}}}}}{{\delta {g^{\mu \nu }}}}=[\phi f(\eta)+\phi-\xi] R_{\mu\nu}- \nabla_\mu \nabla_\nu [\phi f(\eta)+\phi-\xi]
        + g_{\mu\nu}\Box [\phi f(\eta)+\phi-\xi] \nonumber\\
        - \partial _\mu \xi \partial _\nu \eta
         - \frac{w}{\phi }\partial_\mu \phi \partial_\nu \phi.
    \end{eqnarray}
Now, let us consider the flat FRW metric
    \begin{eqnarray}
        ds^2=-dt^2+a(t)^2 \delta _{ij}dx^i dx^j,
    \end{eqnarray}
 and assume that the scalar fields $\phi, \xi$ and $\eta$ only depend on time.
Then, Eq.~(\ref{gravitational equation1}) reduces to
    \begin{eqnarray}
      -6H(H+\frac{d}{dt})\left[\left(f(\eta)+1\right)\phi-\xi\right]
      +\dot{\eta}\dot{\xi}+\frac{w\dot{\phi}}{\phi}+2\kappa^2 \rho&=&0,
       \label{gravitational equation2}\\
       (2\frac{d^2}{dt^2}+4H\frac{d}{dt}+6H^2+4\dot{H})\left[\left(f(\eta)+1\right)\phi-\xi\right]
       +\dot{\eta}\dot{\xi}+\frac{w\dot{\phi}}{\phi}+2\kappa^2 p&=&0,
       \label{gravitational equation3}
    \end{eqnarray}
and  Eqs.~(\ref{equation phi1}-\ref{equation xi1}) read
    \begin{eqnarray}
       6(2H^2+\dot{H})(f(\eta)+1)-6wH\frac{\dot{\phi}}{\phi}
       +w\frac{\dot{\phi}^2-2\phi\ddot{\phi}}{\phi^2}&=&0,
       \label{equation phi2}\\
       6(2H^2+\dot{H})\phi f'(\eta)-3H\dot{\xi}-\ddot{\xi}&=&0,
       \label{equation xi2}\\
       \ddot{\eta}+6\dot{H}+3H(4H+\dot{\eta})&=&0.
       \label{equation eta2}
    \end{eqnarray}

\subsection{De Sitter solutions}
We first focuss on the de Sitter solution with a perfect liquid and assume that
$H(t)=H_0$. The energy-momentum tensor is given by $T_{\mu\nu}=pg_{\mu\nu}+(p+\rho)U_{\mu}U_{\nu}$,
where the equation of state is $\rho=\omega p$. Thus, $\rho=\rho_0 \text{e}^{-3(\omega+1)H_0 t}$.
Then Eq.~(\ref{equation xi2}) can be solved as
    \begin{eqnarray}
        \eta=-4H_0 t -\eta_0 \text{e}^{-3H_0 t}+\eta_1.
    \end{eqnarray}
For simplicity, we consider the special solution $\eta=-4H_0 t$. Note that Eqs.~(\ref{gravitational equation2})-(\ref{equation phi2}) are not independent and, in fact, we just have three  equations for the three variables $\phi$, $\eta$ and $f(\eta)$. Substituting Eq.~(\ref{equation xi2}) into the sum of Eq.~(\ref{gravitational equation2}) and Eq.~(\ref{gravitational equation3}), we get
    \begin{eqnarray}
       (\frac{d^2}{dt^2}-H_0 \frac{d}{dt})\left[\left(f(\eta)+1\right)\phi\right]
       -12H_{0}^{2}\phi f'(\eta)+\frac{w\dot{\phi}}{\phi}+\kappa^2 (p+\rho)&=&0.
       \label{gravitational equation4}
    \end{eqnarray}
The system is now determined by Eqs.~(\ref{equation phi2}) and (\ref{gravitational equation4}) of
$\phi$ and $f(\eta)$. Both of them are second-order non-linear equations. For the vacuum case,
if we consider the late-time acceleration, which implies the scalar curvature $R=12H_{0}^2<<1$,
we can ignore the terms of order $H_{0}^2$ in Eqs.~(\ref{equation phi2}) and (\ref{gravitational equation4}). Then, Eq.~(\ref{equation phi2}) can be solved, approximately, as
    \begin{eqnarray}
        \phi(t)=\phi_0\text{e}^{-H_{0}t}(\text{e}^{\frac{H_{0}t}{2}}-c_1)^2,
    \label{ds phi}
    \end{eqnarray}
and we can consider $c_1=0$ for simplicity. With Eq.~(\ref{ds phi}), Eq.~(\ref{gravitational equation4}) is solved too, to yield
    \begin{eqnarray}
    f(\eta)=-w(\frac{3}{4}\text{e}^{-\frac{\eta}{8}}+\frac{1}{2})
    +\frac{\text{e}^{-\frac{\eta}{4}}}{8\sqrt{1-\text{e}^{-\frac{\eta}{8}}}}
    \left[c_{2}
    -3w\text{ln}(1-2\text{e}^{\frac{\eta}{8}}
    -2\text{e}^{\frac{\eta}{16}}\sqrt{\text{e}^{\frac{\eta}{8}}-1})\right]-1.
    \label{ds f}
    \end{eqnarray}

\subsection{Power scale factor solutions}
In this section we assume that the scale factor is of the form $a(t)=a_0 t^n$ and the Hubble parameter
$H(t)=\frac{n}{t}$. Eq.~(\ref{equation eta2}) can then be solved as
    \begin{eqnarray}
        \eta(t)=\begin{cases}
            \frac{6n(2n-1)}{1-3n}\text{ln}t+\eta_0 t^{1-3n}+\eta_1;& \text{for}~~ n\neq \frac{1}{3}\\
            \frac{1}{3}(\text{ln}t)^2+\eta_0 \text{ln}t +\eta_1;& \text{for}~~ n= \frac{1}{3}.
    \end{cases}
    \end{eqnarray}
Let us consider the case $n=\frac{1}{2}$. In such case, $\ddot{a}(t)<0$, and we can have a decelerating expanding solution.
 With $H(t)=\frac{1}{2t}$ and $\eta(t)=\eta_0 t^{-\frac{1}{2}}+\eta_1$, Eqs.~(\ref{equation phi2}) and (\ref{equation xi2}) have the
solution
\begin{eqnarray}
   \phi(t) &=& \frac{\phi_0}{t} + \frac{2\sqrt{\phi_0 \phi_1}}{\sqrt{t}}+\phi_1,
   \label{phi}\\
   \xi(t) &=& \frac{\xi_0}{\sqrt{t}}+\xi_1.
   \label{xi}
\end{eqnarray}
For simplicity, we set $\phi_1=\xi_1=\eta_1=0$, and Eq.~(\ref{gravitational equation3})
(or equivalently Eq.~(\ref{gravitational equation4}) has the solution
\begin{eqnarray}
   f(\eta)=\frac{f_0 \eta_0}{\eta}+\frac{4\kappa^2\rho_0}{\phi_0 (6-9\omega)}
   \left(\frac{\eta_0}{\eta}\right)^{\frac{3}{2}(1-\omega)}-\frac{\eta_0 \xi_0}{6\phi_0}-\frac{2w}{3}-1.
   \label{f}
\end{eqnarray}


\section{Mimetic Gauss-Bonnet gravity}
Now we turn to  mimetic Gauss-Bonnet gravity. The action of mimetic Gauss-Bonnet gravity is
    \begin{eqnarray}
        S=\int d^4x\sqrt{-g(\hat{g}_{\mu\nu},\phi)} \{ \frac{1}{2\kappa^2}[\phi(R(\hat{g}_{\mu\nu},\phi)+b\mathcal{G}
        -\frac{w}{\phi } \nabla _{\mu }\phi \nabla ^{\mu }\phi]
       -V(\varphi)\} + S_M
        \label{action nbd0}
    \end{eqnarray}
Eq. (\ref{memetic metric}) shows the gauge freedom is constrained by
    \begin{eqnarray}
        g(\hat{g}_{\mu\nu},\phi)^{\rho\sigma} \pd_{\rho}\phi\pd_{\sigma}\phi \hat{g}_{\mu\nu}.
        \label{mimetic gauge}
    \end{eqnarray}
For convenience, we impose the constrain (\ref{mimetic gauge}) by introducing the Lagrangian multiplier
$\lambda$, and the action (\ref{action nbd0}) turns into
    \begin{eqnarray}
        S=\int d^4x\sqrt{-g} \{ \frac{1}{2\kappa^2}[\phi(R+b\mathcal{G}
        -\frac{w}{\phi } \nabla _{\mu }\phi \nabla ^{\mu }\phi]
        +\lambda(\nabla _{\mu }\varphi \nabla ^{\mu }\varphi+1)-V(\varphi)\} + S_M,
        \label{action nbd1}
    \end{eqnarray}
where the Gauss-Bonnet term $\mathcal{G}$ is
    \begin{eqnarray}
        \mathcal{G}= R^2-4R_{\mu\nu}R^{\mu\nu}+R_{\mu\nu\lambda\rho}R^{\mu\nu\lambda\rho}.
    \end{eqnarray}
By varying now with respect to $g_{\mu\nu}$, we have the EoM of the gravitational field
    \begin{eqnarray}
        \phi(R_{\mu\nu}-\frac{1}{2}R g_{\mu\nu})
        -\frac{\omega}{\phi}(\nabla_{\mu}\phi \nabla_{\mu}\phi-\frac{1}{2}g_{\mu\nu}\nabla _{\mu }\phi \nabla ^{\mu }\phi)
        +(g_{\mu\nu}\Box-\nabla_{\mu}\nabla_{\nu})\phi+bH_{\mu\nu}\\
        =\kappa^2 T_{\mu\nu}
        +\kappa^2 g_{\mu\nu}[\lambda(\nabla_{\mu }\varphi \nabla^{\mu}\varphi+1)-V(\varphi)]
        -2\kappa^2 \lambda \nabla_{\mu}\varphi \nabla_{\nu}\varphi,
    \end{eqnarray}
where $H_{\mu\nu}$ is given by
    \begin{eqnarray}
        H_{\mu\nu}=2R(g_{\mu\nu}\Box-\nabla_{\mu}\nabla_{\nu})\phi
        +4R_{\mu}^{\alpha}\nabla_{\alpha}\nabla_{\nu}\phi
        +4R_{\nu}^{\alpha}\nabla_{\alpha}\nabla_{\mu}\phi\\
        -4R_{\mu\nu}\Box\phi
        -4g_{\mu\nu}R^{\alpha\beta}\nabla_{\alpha}\phi \nabla_{\beta}\phi
        +4R_{\alpha\mu\beta\nu}\nabla^{\alpha}\phi \nabla^{\beta}\phi.
    \end{eqnarray}
Variation with respect to $\phi$ yields the kinematical wave equation
    \begin{eqnarray}
       \frac{2w}{\phi}\Box\phi + R-\frac{w}{\phi^2}\nabla_{\mu}\phi\nabla^{\mu}\phi+b\mathcal{G}=0,
    \end{eqnarray}
while variation with respect to $\varphi$ gives
    \begin{eqnarray}
       2\nabla^{\mu}(\lambda\nabla_{\mu}\varphi)+\frac{\pd V}{\pd \varphi}=0,
    \end{eqnarray}
and variation with respect to $\lambda$ yields
    \begin{eqnarray}
       \nabla_{\mu}\varphi\nabla^{\mu}\varphi+1=0,
       \label{eom varphi}
    \end{eqnarray}
being the metric
    \begin{eqnarray}
        ds^2=-dt^2+a(t)^2 \delta _{ij}dx^i dx^j,
        \label{metric}
    \end{eqnarray}
Then
    \begin{eqnarray}
       -\kappa^2 V(\varphi) +3H^2\phi+3H\dot{\phi}+12bH^3\dot{\phi}
       -\frac{w\dot{\phi}^2}{2\phi}+\kappa^2\lambda(1+\dot{\varphi})&=&0
       \label{eom grav1}\\
       \kappa^2 V(\varphi) -3H^2\phi-2\dot{H}\phi-2H\dot{\phi}-8bH^3\dot{\phi}
       -8bH\dot{H}\dot{\phi} \nonumber\\
       -\frac{w\dot{\phi}^2}{2\phi}+\kappa^2\lambda(-1+\dot{\varphi}^2)
       -\ddot{\phi}-4bH^2\ddot{\phi}&=&0
       \label{eom grav2}\\
       24bH^4+6\dot{H}+12H^2(1+2b\dot{H})-\frac{6wH\dot{\phi}}{\phi}
       +\frac{w(\dot{\phi}^2-2\phi\ddot{\phi})}{\phi^2}=0
       \label{eom phi}\\
       V'(\varphi)-2(3H\lambda \dot{\varphi}
       +\dot{\lambda}\dot{\varphi}+\lambda\ddot{\varphi})=0.
       \label{eom lambda}
    \end{eqnarray}
 Eq. (\ref{eom varphi}) reads actually
    \begin{eqnarray}
       -\dot{\varphi}^2+1=0,
    \end{eqnarray}
thus $\varphi$ can be identified with the cosmological time: $\varphi=t$.
\subsection{dS Solutions}
First, we consider the dS solution, for which we have $H(t)=H_0$. The general solution of Eq. (\ref{eom phi}) is
    \begin{eqnarray}
       \phi(t)=\phi_0 \text{e}^{-3H_0 t} \text{cosh}^2(kt-c_1),
       \label{sol phi1},
    \end{eqnarray}
where $\phi_0$ and $c_1$ are integration constants, and
    \begin{eqnarray}
       k=\frac{\sqrt{3}H_0}{2}\sqrt{\frac{4+8bH_{0}^2+3w}{w}}.
       \label{sol phi2}
    \end{eqnarray}
For simplicity, we assume that $c_1=0$. With Eqs. (\ref{sol phi1}) and (\ref{sol phi1}), we can
solve $V(\phi)$ as a function of $\lambda$, and substituting this into Eq. (\ref{eom lambda}),
we finally have the solutions
    \begin{eqnarray}
        \lambda(t)=\frac{1}{2}(1+\omega)\text{e}^{-3(1+\omega)H_{0}t}
        +\frac{6b\phi_0 H_0^{4}}{\kappa^2}\text{e}^{-3H_0 t}
        +\text{e}^{-3H_0 t}
        \left[ \lambda_1 \text{cosh}(2kt)+\lambda_2  \text{sinh}(2kt)\right],
    \end{eqnarray}
where
    \begin{eqnarray}
        \lambda_1&=&\frac{3H_{0}^2\phi_0}{4w\kappa^2}
        \left(4+32b^2H_{0}^4+11w+6w^2+24bH_{0}^2+36wbH_{0}^2\right)\\
          \lambda_2&=&\frac{H_{0}^2\phi_0}{4\kappa^2}
        \left[24b^2H_{0}^2-2k(7+28b^2H_{0}^2+6w)\right].
    \end{eqnarray}
Finally, the scalar potential $V$ is given by
    \begin{eqnarray}
        V(t)=w\text{e}^{-3(1+\omega)H_{0}t}
        +\frac{\phi_0 H_{0}^2}{2w\kappa^2}\text{e}^{-3H_0 t}
        [3(1+4bH_{0}^2+w)(4+8bH_{0}^2+3w)\text{cosh}(2kt)\nonumber\\
        -\sqrt{3w(4+8bH_{0}^2+3w)}(4+16bH_{0}^2+3w)\text{sinh}(2kt)].
    \end{eqnarray}

\subsection{Power scale factor Solutions}
We now consider the power scale factor solution, $a(t)=a_0 t^n$. The EoS for matter
is $p=\omega \rho$, and we have $\rho(t)=a(t)^{-3-3\omega}$.
Eq. (\ref{eom phi}) has the solution
    \begin{eqnarray}
       \phi(t)=\phi_0 t^{1-3 n} I_{\alpha }\left(\frac{k}{t}\right){}^2
       \label{sol phi1}
    \end{eqnarray}
Here we have assume one of the integral constant to be $0$ and the constants $\alpha$ and $k$ are given by
    \begin{eqnarray}
       \alpha &=& \frac{\sqrt{3 n^2 (3 w+4)-6 n (w+1)+w}}{2 \sqrt{w}},\\
       k&=&\frac{\sqrt{6} \sqrt{b (n-1) n^3}}{\sqrt{w}},
    \end{eqnarray}
respectively. Then, we substitute them into Eqs. (\ref{eom grav1}) and (\ref{eom lambda}) and obtain the following solutions for $\lambda(t)$ and $V(t)$:
    \begin{eqnarray}
       \lambda(t) &=&\frac{1}{12 \kappa ^2 n}t^{-3 n-5} \left(t^n\right)^{-3 \omega }
       \{6\kappa ^2 n t^5 (\omega +1)\nonumber\\
       &&+\text{$\phi $0} +\left(t^n\right)^{3 \omega }[12 k^2 n \left(4 b n^2+t^2 (2 w+1)\right) I_{\alpha -1}\left(\frac{k}{t}\right){}^2\nonumber\\
       &&+2 k t(48 b n^3 (-\alpha +3 n+1)+4 k^2 w\nonumber\\
       &&+t^2 \left(3 n^2 (9 w+10)-6 n (2 \alpha +4 \alpha  w+w)+\left(4 \alpha ^2-1\right) w\right))I_{\alpha -1}\left(\frac{k}{t}\right) I_{\alpha }\left(\frac{k}{t}\right)\nonumber\\
       &&24 b n^3 \left(2 k^2+t^2 (-2 \alpha +3 n-1) (-2 \alpha +3 n+3)\right)\nonumber\\
       &&+t^2\left(4 k^2 (3 n (w+1)-(2 \alpha +1) w)\right.\nonumber\\
       &&+t^2(-2 \alpha +3 n+1)\left(3 n^2 (3 w+4)+w\right.\nonumber\\
       &&+4 \alpha  (\alpha +1) w-6 (2 \alpha +1) n (w+1))))I_{\alpha }\left(\frac{k}{t}\right){}^2
       ]\}
	\end{eqnarray}
    \begin{eqnarray}
       V(t)&=&-\frac{1}{6 \kappa ^2}t^{-3 n-5} \left(t^n\right)^{-3 \omega } \nonumber\\
       &&\left(3 t \left(\text{$\phi $0} \left(t^n\right)^{3 \omega } \left(4 k I_{\alpha }\left(\frac{k}{t}\right) I_{\alpha -1}\left(\frac{k}{t}\right) \left(12 b n^3+t^2 (3 n (w+1)-(2 \alpha +1) w)\right) \right.\right.\right.
       \nonumber \\
        &&\left.\left.\left.+t I_{\alpha }\left(\frac{k}{t}\right){}^2 \left(24 b n^3 (-2 \alpha +3 n-1)+t^2 \left(3 n^2 (3 w+4)-6 (2 \alpha +1) n (w+1)+4 \alpha  (\alpha +1) w+w\right)\right)
        \right.\right.\right.
       \nonumber \\
        &&\left.\left.\left.
       +4 k^2 t w I_{\alpha -1}\left(\frac{k}{t}\right){}^2\right)+2 \kappa ^2 t^4\right) \right.
       \nonumber \\
        &&\left.-\frac{1}{n}\text{$\phi $0} \left(t^n\right)^{3 \omega } \left(12 k^2 n \left(4 b n^2+t^2 (2 w+1)\right) I_{\alpha -1}\left(\frac{k}{t}\right){}^2\right.\right.
       \nonumber \\
        &&\left.\left.+2 k t I_{\alpha }\left(\frac{k}{t}\right) I_{\alpha -1}\right.\right.
       \nonumber \\
        &&\left.\left.\left(\frac{k}{t}\right) \left(48 b n^3 (-\alpha +3 n+1)+4 k^2 w+t^2 \left(3 n^2 (9 w+10)-6 n (2 \alpha +4 \alpha  w+w)+\left(4 \alpha ^2-1\right) w\right)\right)\right.\right.
       \nonumber \\
        &&\left.\left.+I_{\alpha }\left(\frac{k}{t}\right){}^2 \left(24 b n^3 \left(2 k^2+t^2 (-2 \alpha +3 n-1) (-2 \alpha +3 n+3)\right)\right.\right.\right.
       \nonumber \\
        &&\left.\left.\left.+t^2 \left(4 k^2 (3 n (w+1)-(2 \alpha +1) w)+t^2 (-2 \alpha +3 n+1) \left(3 n^2 (3 w+4)-6 (2 \alpha +1) n (w+1)\right.\right.\right.\right.\right.
       \nonumber \\
        &&\left.\left.\left.\left.\left.+4 \alpha  (\alpha +1) w+w\right)\right)\right)\right)+6 \kappa ^2 n t^5 (\omega +1)\right)
    \end{eqnarray}

\section{Conclusion}
In this paper, we have obtained new cosmological solutions of non-local Brans-Dicke gravity and mimetic Gauss-Bonnet gravity. Similarly to what happened with the introduction of extended GR theories, extensions of BD gravity naturally come into play here. To this we have here added the possibility, as discussed in the paper, of the theory being non-local and have obtained cosmological solutions for a BD non-local model. We have first realized the de Sitter solutions corresponding to the accelerating expanding periods and, later, the power scale factor cosmology, corresponding to a perfect fluid.

 For the case of non-local Brans-Dicke gravity, the dS solution of the late-time acceleration has been found by imposing the scalar curvature to be small, $R<<1$. We have also obtained the power scale factor solution of $a(t)=a_0 t^{\frac{1}{2}}$. It corresponds to a  decelerating expanding solution. In the case of  mimetic Gauss-Bonnet gravity, the power scale factor solution is given by $a(t)=a_0 t^{n}$, with $n$ an arbitrary constant.

\section*{Acknowledgement}

Y.Z. would like to acknowledge the support of a scholarship granted by the Chinese Scholarship Council (CSC).
E.E.  was supported by  MINECO (Spain), Project FIS2013-44881, and by I-LINK 1019 (CSIC).


\end{document}